\documentclass[12pt]{article}

\usepackage{algorithm}				% algorithms
\usepackage[noend]{algpseudocode}	% algorithms
\usepackage{amsmath}				% math package
\usepackage{amssymb}				% math package
\usepackage{amsthm}				% math package
\usepackage{caption}
\usepackage{subcaption}
\usepackage{color}
\usepackage[margin=1in]{geometry}
\usepackage[hidelinks]{hyperref}
\usepackage{tikz}
\usepackage{titling}

\usetikzlibrary{matrix, shapes}

\usepackage{adjustbox}			% rotatebox
\usepackage{complexity}			% complexity classes
\usepackage{stmaryrd}			% concatenation symbols

\theoremstyle{plain}				% AMS theorem styles
\newtheorem{theorem}{Theorem}
\newtheorem{lemma}[theorem]{Lemma}
\newtheorem{proposition}[theorem]{Proposition}
\newtheorem{corollary}[theorem]{Corollary}

\theoremstyle{definition}			% AMS definition styles
\newtheorem{definition}[theorem]{Definition}

\theoremstyle{remark}			% AMS remark styles
\newtheorem*{remark}{Remark}

\newcommand{\REC}{\textsf{REC}}
\newcommand{\TDFA}{\textsf{2DFA-4W}}
\newcommand{\TDFAOS}{\textsf{2DFA-4W-1$\Sigma$}}

\newcommand{\TDFATW}{\textsf{2DFA-3W}}
\newcommand{\TDFATWOS}{\textsf{2DFA-3W-1$\Sigma$}}
\newcommand{\TDFATWOW}{\textsf{2DFA-2W}}

\newcommand{\TNFA}{\textsf{2NFA-4W}}
\newcommand{\TNFAOS}{\textsf{2NFA-4W-1$\Sigma$}}

\newcommand{\TNFATW}{\textsf{2NFA-3W}}
\newcommand{\TNFATWOS}{\textsf{2NFA-3W-1$\Sigma$}}
\newcommand{\TNFATWOW}{\textsf{2NFA-2W}}

\DeclareMathOperator{\proj}{pr}
\DeclareMathOperator{\projrow}{R}
\DeclareMathOperator{\projcol}{C}
\newcommand{\rowp}[1]{\ensuremath{\proj_{\projrow}(#1)}}
\newcommand{\colp}[1]{\ensuremath{\proj_{\projcol}(#1)}}

\DeclareMathOperator{\rowsubword}{rso}

\thanksmarkseries{alph}

\title{Recognition and Complexity Results for Projection Languages of Two-Dimensional Automata}
\author{Taylor J. Smith \thanks{School of Computing, Queen's University, Kingston, Ontario, Canada. Email: \texttt{\{tsmith,ksalomaa\}@cs.queensu.ca}.} \and Kai Salomaa \thanksmark{1}}
\date{\today}

\begin{document}

%%%%%%

\maketitle

\begin{abstract}
The row projection (resp., column projection) of a two-dimensional language $L$ is the one-dimensional language consisting of all first rows (resp., first columns) of each two-dimensional word in $L$. The operation of row projection has previously been studied under the name ``frontier language", and previous work has focused on one- and two-dimensional language classes.

In this paper, we study projections of languages recognized by various two-dimensional automaton classes. We show that both the row and column projections of languages recognized by (four-way) two-dimensional automata are exactly context-sensitive. We also show that the column projections of languages recognized by unary three-way two-dimensional automata can be recognized using nondeterministic logspace. Finally, we study the state complexity of projection languages for two-way two-dimensional automata, focusing on the language operations of union and diagonal concatenation.

\medskip


\medskip

\noindent\textit{MSC2020 classes:} 68Q45 (primary); 68Q15, 68Q19 (secondary).
\end{abstract}

%%%%%%

\section{Introduction}\label{sec:introduction}

A two-dimensional word is a generalization of the notion of a word from a one-dimensional string to an array or matrix of symbols. Two-dimensional words are used as the input to two-dimensional automata, whose input heads move through the input word in a variety of ways, depending on the model.

We may define special projection operations on two-dimensional words that produce either the first row or the first column of the given word. In this way, a projection can be thought of as a conversion from a two-dimensional word to a one-dimensional word. Note that projection operations are lossy (i.e., all but the first row/column of the two-dimensional word is lost when a projection operation is applied).

The row projection operation has been studied in the past \cite{Anselmo2011ClassificationTilingRecognizable, Latteux1997ContextSensitiveRecognizable}, with a particular focus on formal language theory. (We summarize previous results in Section~\ref{subsec:prevwork}.) However, no work has yet been done on investigating projections of languages recognized by various two-dimensional automaton models.

Our results are as follows. We show that both the row and column projections of languages recognized by (four-way) two-dimensional automata are exactly context-sensitive. We also show that the column projections of languages recognized by unary three-way two-dimensional automata belong to the class $\NSPACE(O(\log(n)))$. Finally, we study the state complexity of projection languages, focusing on the state complexity of union and diagonal concatenation for projections of languages recognized by two-way two-dimensional automata.

%%%%%%

\section{Preliminaries}\label{sec:prelim}

A two-dimensional word is a matrix of symbols from some alphabet $\Sigma$. If a two-dimensional word $w$ has $m$ rows and $n$ columns, then we say that $w$ is of dimension $m \times n$. A two-dimensional language consists of two-dimensional words. There exist two special languages in two dimensions: $\Sigma^{m \times n}$ consists of all words of dimension $m \times n$ for some fixed $m, n \geq 1$, and $\Sigma^{**}$ consists of all two-dimensional words.

The row projection (resp., column projection) of a two-dimensional language $L$ is the one-dimensional language consisting of the first rows (resp., first columns) of all two-dimensional words in $L$. We formalize these definitions in terms of individual two-dimensional words. In the following pair of definitions, we assume we have an $m \times n$ two-dimensional word
\begin{equation*}
w = 
\begin{bmatrix}
a_{1,1} 	& \cdots 	& a_{1,n} \\
\vdots	& \ddots	& \vdots \\
a_{m,1}	& \cdots	& a_{m,n}
\end{bmatrix}
.
\end{equation*}

\begin{definition}[Row projection]
Given a two-dimensional word $w \in \Sigma^{m \times n}$, the row projection of $w$ is the one-dimensional word
\begin{equation*}
\rowp{w} = a_{1,1} a_{1,2} \cdots a_{1,n},
\end{equation*}
where $a_{1,1}, \dots, a_{1,n} \in \Sigma$. The row projection of a two-dimensional language $L$, denoted $\rowp{L}$, is produced by taking the row projections of all words $w \in L$.
\end{definition}

\begin{definition}[Column projection]
Given a two-dimensional word $w \in \Sigma^{m \times n}$, the column projection of $w$ is the one-dimensional word
\begin{equation*}
\colp{w} = a_{1,1} a_{2,1} \cdots a_{m,1},
\end{equation*}
where $a_{1,1}, \dots, a_{m,1} \in \Sigma$. The column projection of a two-dimensional language $L$, denoted $\colp{L}$, is produced by taking the column projections of all words $w \in L$.
\end{definition}

Note that one may view the column projection operation as taking the ``transpose" of the first column of a two-dimensional word in order to produce a one-dimensional string. The row projection operation has been considered in previous papers, where it was called the ``frontier" of a word or language \cite{Latteux1997ContextSensitiveRecognizable}.

Two-dimensional words are used as the input to two-dimensional automata. When we provide such a word as input, we surround the outer border of the word with a special boundary symbol \#. (For example, the upper-left boundary symbol is at position $(0,0)$ and the lower-right boundary symbol is at position $(m+1, n+1)$ in the word.) The boundary symbol prevents the input head of the automaton from leaving the input word.

The formal definition of a two-dimensional automaton is as follows:

\begin{definition}[Two-dimensional automaton]\label{def:2DFA}
A two-dimensional 
automaton is a tuple $(Q, \Sigma, \delta, q_{0}, q_{\rm accept})$, where $Q$ is a finite set of states, $\Sigma$ is the input alphabet (with $\# \not\in \Sigma$ acting as a boundary symbol), $\delta: (Q \setminus \{q_{\rm accept}\}) \times (\Sigma \cup \{\#\}) \to Q \times \{U, D, L, R\}$ is the partial transition function, and $q_{0}, q_{\rm accept} \in Q$ are the initial and accepting states, respectively.
\end{definition}

The specific model in Definition~\ref{def:2DFA} is sometimes referred to as a ``four-way two-dimensional automaton". In this paper, we also consider three-way and two-way variants of two-dimensional automata. In the three-way case, the transition function is restricted to use only the directions $\{D, L, R\}$. Likewise, in the two-way case, the transition function uses only the directions $\{D, R\}$. We may optionally include a direction $N$, which corresponds to ``no move" and does not change the recognition power of the model. We abbreviate each automaton model as \textsf{2(D/N)FA-kW(-1$\Sigma$)}, where \textsf{D/N} denotes deterministic/nondeterministic, \textsf{k} denotes the directions of movement, and \textsf{1$\Sigma$} denotes a unary alphabet.
In later sections, we will use the notation $L_{\textsf{C}}$ to denote the set of languages recognized by some automaton model \textsf{C}.

%%%

\subsection{Previous Work}\label{subsec:prevwork}

A number of survey articles and other works have been written about both two-dimensional languages \cite{GiammarresiRestivo19972DLanguages, Morita20042DLanguages} and two-dimensional automaton models \cite{Inoue19912DAutomataSurvey, Rosenfeld1979PictureLanguages, Smith2019TwoDimensionalAutomata}. Previous work on projection operations has taken two perspectives: language-theoretic and automata-theoretic.

\paragraph{Language-theoretic.}
One of the earliest results on two-dimensional row projection, due to Latteux and Simplot \cite{Latteux1997ContextSensitiveRecognizable}, showed that a one-dimensional language $F$ is context-sensitive if and only if there exists a two-dimensional language $L \in \REC$ such that $F = \rowp{L}$. The class $\REC$ denotes the class of tiling-recognizable two-dimensional languages, or languages whose words can be defined by a finite set of $2 \times 2$ tiles \cite{GiammarresiRestivo1992RecognizablePictureLanguages}.

Anselmo et al.\ \cite{Anselmo2011ClassificationTilingRecognizable} later extended this direction of research to give equivalent characterizations for unambiguous and deterministic context-sensitive one-dimensional languages; namely, $F$ is unambiguous (resp., deterministic) context-sensitive if and only if there exists $L \in \textsf{UREC}$ (resp., $L \in \textsf{Row-UREC}_{t}$) such that $F = \rowp{L}$. The classes \textsf{UREC} and $\textsf{Row-UREC}_{t}$ are subclasses of $\REC$, where \textsf{UREC} consists of languages defined by an unambiguous tiling system \cite{GiammarresiRestivo1992RecognizablePictureLanguages} and $\textsf{Row-UREC}_{t}$ consists of languages that are ``top-to-bottom row-unambiguous"; Anselmo et al.\ give a formal definition of the class $\textsf{Row-UREC}_{t}$ in an earlier paper \cite{Anselmo2010DeterministicUnambiguousFamilies}.

Some classes smaller than $\textsf{Row-UREC}_{t}$ (namely, the class of deterministic recognizable languages $\textsf{DREC}$ \cite{Anselmo2010DeterministicUnambiguousFamilies}) have no known characterization in terms of one-dimensional language classes.

\paragraph{Automata-theoretic.}
A (four-way) two-dimensional automaton can recognize whether or not an input word has either an exponential or a doubly-exponential side length \cite{KariMoore2004RectanglesAndSquares}. It is well-known that the language of unary strings of exponential length is context-sensitive but not context-free \cite{HopcroftUllman1979IntroAutomataTheory}. This fact implies that, if $L$ is a language recognized by a four-way two-dimensional automaton, then both $\rowp{L}$ and $\colp{L}$ may be non-context-free, even in the unary case.

Restricting ourselves to the three-way model, we obtain results that differ based on the projection operation under consideration. Let $L$ be a unary language. If $L$ is recognized by a nondeterministic three-way two-dimensional automaton, then $\rowp{L}$ is regular. On the other hand, if $L$ is recognized by a deterministic three-way two-dimensional automaton, then $\colp{L}$ need not be regular \cite{SmithSalomaa20192D3WProperties}. These results apply also for general alphabets. We can improve the bound by showing that $\colp{L}$ may be non-context-free for three-way two-dimensional automata, since the language $L_{\text{composite}}$ used in the proof of the non-regularity result is context-sensitive in both the unary and general-alphabet cases \cite{Hartmanis1968RecognitionPrimesAutomata, Salomaa1969TheoryOfAutomata, Salomaa1973FormalLanguages}.

Finally, for the two-way model, we know that if any language $L$ is recognized by a nondeterministic two-way two-dimensional automaton, then both $\rowp{L}$ and $\colp{L}$ are regular \cite{SmithSalomaa20192D3WProperties}. This applies also to deterministic and unary two-way two-dimensional automata.

%%%%%%

\section{Recognition Power and Space Complexity}\label{sec:recognition}

Before we proceed further, we recall a few elementary definitions. These definitions may be found in any standard textbook on the theory of computation; e.g., Sipser~\cite{Sipser1997Computation}.

Recall that a linear-bounded automaton is a nondeterministic Turing machine whose computation is restricted only to the cells of its input tape that originally contained input symbols. A configuration of a linear-bounded automaton $\mathcal{M}$ is a sequence of tape symbols of $\mathcal{M}$, where the currently-scanned symbol is distinguished by adding the current state $q$ as a subscript to the symbol. An accepting computation history of $\mathcal{M}$ on an input string $w$ is a sequence of configurations $C_{0}, C_{1}, \dots, C_{k}$ that $\mathcal{M}$ enters as it performs its computation on $w$, where $C_{0}$ is the initial configuration of $\mathcal{M}$ on $w$, $C_{i+1}$ is obtained from $C_{i}$ in one computation step of $\mathcal{M}$ for all $0 \leq i \leq k - 1$, and $C_{k}$ is an accepting configuration. 
Finally, a computation table of $\mathcal{M}$ on an input word $w$ is a two-dimensional word where the rows of the word are configurations $C_{0}, C_{1}, \dots, C_{k}$ appearing in the computation history of $\mathcal{M}$ on $w$.

From previous work, we know that $\rowp{L}$ is context-sensitive when $L \in \REC$ \cite{Latteux1997ContextSensitiveRecognizable}. It is known that $L_{\TDFA} \subset L_{\TNFA} \subseteq \REC$ \cite{BlumHewitt19672DAutomata, Inoue1992CharacterizationRecognizable}, so $\rowp{L}$ is also context-sensitive when $L \in L_{\TDFA}$. The following theorem proves the other direction of this inclusion.

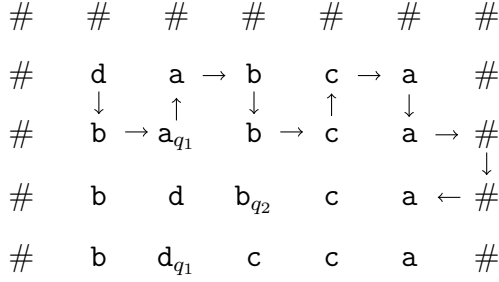
\begin{figure}
\centering
\begin{tikzpicture}
\matrix[matrix of nodes,nodes={inner sep=4pt,text width=.75cm,align=center,minimum height=.75cm}]{
\node(a0){\#};	& \node(a1){\#};			& \node(a2){\#};					& \node(a3){\#};					& \node(a4){\#};			& \node(a5){\#};			& \node(a6){\#}; \\
\node(a00){\#};	& \node(a01){\texttt{d}};	& \node(a02){\texttt{a}};			& \node(a03){\texttt{b}};			& \node(a04){\texttt{c}};	& \node(a05){\texttt{a}};	& \node(a06){\#}; \\
\node(a10){\#};	& \node(a11){\texttt{b}};	& \node(a12){\texttt{a}$_{q_{1}}$};	& \node(a13){\texttt{b}};			& \node(a14){\texttt{c}};	& \node(a15){\texttt{a}};	& \node(a16){\#}; \\
\node(a20){\#};	& \node(a21){\texttt{b}};	& \node(a22){\texttt{d}};			& \node(a23){\texttt{b}$_{q_{2}}$};	& \node(a24){\texttt{c}};	& \node(a25){\texttt{a}};	& \node(a26){\#}; \\
\node(a30){\#};	& \node(a31){\texttt{b}};	& \node(a32){\texttt{d}$_{q_{1}}$};	& \node(a33){\texttt{c}};			& \node(a34){\texttt{c}};	& \node(a35){\texttt{a}};	& \node(a36){\#}; \\
};

\draw[->] (a01)+(0mm, -2.5mm) -- +(0mm, -5.5mm);
\draw[->] (a11)+(3.5mm, 0mm) -- +(6.5mm, 0mm);
\draw[->] (a12)+(0mm, 2.5mm) -- +(0mm, 5.5mm);
\draw[->] (a02)+(3.5mm, 0mm) -- +(6.5mm, 0mm);
\draw[->] (a03)+(0mm, -2.5mm) -- +(0mm, -5.5mm);
\draw[->] (a13)+(3.5mm, 0mm) -- +(6.5mm, 0mm);
\draw[->] (a14)+(0mm, 2.5mm) -- +(0mm, 5.5mm);
\draw[->] (a04)+(3.5mm, 0mm) -- +(6.5mm, 0mm);
\draw[->] (a05)+(0mm, -2.5mm) -- +(0mm, -5.5mm);
\draw[->] (a15)+(3.5mm, 0mm) -- +(6.5mm, 0mm);

\draw[->] (a16)+(0mm, -2.5mm) -- +(0mm, -5.5mm);
\draw[->] (a26)+(-3.5mm, 0mm) -- +(-6.5mm, 0mm);
\end{tikzpicture}
\caption{An illustration of the movement of the input head of the automaton $\mathcal{A}$, constructed in Theorem~\ref{thm:CSLin2DFA4W}.}
\label{fig:inputheadmovement}
\end{figure}

\begin{theorem}\label{thm:CSLin2DFA4W}
Let $K$ be a context-sensitive language. Then there exists $L \in L_{\TDFA}$ such that $K = \rowp{L}$.

\begin{proof}
Let $\mathcal{M}$ be a linear-bounded automaton recognizing the language $K$. 
We construct a deterministic four-way two-dimensional automaton $\mathcal{A}$ that checks whether the rows of its input word $w$ are configurations representing an accepting computation history of $\mathcal{M}$\footnote{Since the row projection of $w$ cannot contain state information, an initial configuration of $\mathcal{M}$, $a_{1, q_{0}} a_{2} \dots a_{m}$, is encoded on the first row as the string $a_{1} a_{2} \dots a_{m}$. The computation of $\mathcal{A}$ implicitly assumes that this string represents the initial configuration.}.
Given a two-dimensional word $w$ over an alphabet $\Sigma$ as input, where $w$ is of dimension $m \times n$, $\mathcal{A}$ checks each of the following properties:
\begin{enumerate}
\item The first row of $w$ contains only alphabet symbols from $\Sigma$.
\item The last row of $w$ contains, as a subscript, an accepting state of $\mathcal{M}$.
\item\label{itm:prop3} For each $i$, where $1 \leq i \leq m - 1$, the configuration $C_{i}$ represented by the $(i + 1)$st row can be obtained in one computation step of $\mathcal{M}$ from the preceding configuration $C_{i-1}$ represented by the $i$th row.

For this property, $\mathcal{A}$ must check the following:
	\begin{enumerate}
	\item Any tape symbol in $C_{i}$ different from the corresponding symbol in $C_{i-1}$ was first scanned by the state in $C_{i-1}$; and
	\item Each tape symbol in $C_{i}$ different from the corresponding symbol in $C_{i-1}$ corresponds to one valid computation step of $\mathcal{M}$.
	\end{enumerate}
\end{enumerate}
At the beginning of its computation, $\mathcal{A}$ assumes that $\mathcal{M}$ is scanning the leftmost symbol in the first row of $w$ from its initial state. Then, according to Property~\ref{itm:prop3}, $\mathcal{A}$ checks that the configuration represented by the second row of $w$ can be obtained in one computation step of $\mathcal{M}$ under this assumption.

The automaton $\mathcal{A}$ checks Property~\ref{itm:prop3} in the following way. For each $1 \leq i \leq m - 1$, the input head of $\mathcal{A}$ traverses the $i$th and $(i+1)$st rows in a down-and-up motion from the left boundary to the right boundary. Upon reaching a boundary, the input head returns to the input word and checks the next pair of rows. This traversal  procedure is illustrated in Figure~\ref{fig:inputheadmovement}. Each time the input head moves downward or upward within a column, it either compares the symbols in the $i$th and $(i+1)$st rows to verify that they match or, if it is at a position where the current computation step of $\mathcal{M}$ applies, it checks that the computation step is valid according to the transition relation of $\mathcal{M}$.

If each of the preceding properties holds, then $w$ encodes a computation table corresponding to an accepting computation of $\mathcal{M}$ and $\mathcal{A}$ accepts $w$. 
Since linear-bounded automata recognize all context-sensitive languages, and since $\rowp{L(\mathcal{A})}$ is the input string to $\mathcal{M}$, the result follows.
\end{proof}
\end{theorem}

The proof of Theorem~\ref{thm:CSLin2DFA4W} also works for nondeterministic two-dimensional automata. Moreover, it is straightforward to show that $\colp{L}$ is context-sensitive when $L \in L_{\TDFA}$, and so Theorem~\ref{thm:CSLin2DFA4W} can similarly be adapted to apply to column projection languages. These observations, taken together, lead to the following characterization.

\begin{corollary}\label{cor:4WautomataCSL}
Both the row and column projections of languages recognized by four-way two-dimensional automata consist exactly of the class of context-sensitive languages.
\end{corollary}

%%%

\subsection{Three-Way Two-Dimensional Automata}\label{subsec:recognition3W}

Recall from Section~\ref{subsec:prevwork} that the row projection of any language accepted by a three-way two-dimensional automaton $\mathcal{A}$ is regular. Since $\REG \in \DSPACE(O(1))$ \cite{Shepherdson1959TwoWayReduction}, we immediately get that $\rowp{L(\mathcal{A})} \in \DSPACE(O(1))$ as well.

We further noted in the same section that the column projection of a language in $L_{\TNFATWOS}$ may be non-context-free, depending on the choice of language. Here, we investigate the space complexity of column projection languages for $L_{\TNFATWOS}$.

In what follows, we use the notation $\rowsubword_{i}[r, s]$ to denote the subword occurrence of the $i$th row starting at index $r$ and ending at index $r + s$; that is, a subword of length $s + 1$. Since we are considering unary languages, all symbols of $\rowsubword_{i}[r, s]$ are identical and independent of the value $i$. Thus, by ``subword occurrence", we mean the cells of the $i$th row at indices $r$ through $r + s$ inclusive.

The following technical lemma states that every string $w$ in the column projection of a language in $L_{\TNFATWOS}$ is a projection of a two-dimensional word $z$, where the number of columns of $z$ is at most some constant multiple of the length of $w$. The proof, intuitively speaking, shows that when we have a two-dimensional word containing a large number of columns with no downward moves, we can remove some of these columns and simulate the same computation of the three-way two-dimensional automaton.

\begin{lemma}\label{lem:reduceworddimension}
Let $\mathcal{A}$ be a unary three-way two-dimensional automaton with $k$ states, and consider a word $w \in \colp{L(\mathcal{A})}$. Then there exists a two-dimensional word $z$ with $|w|$ rows and at most $(|w| + 3) \cdot (k^{2^{2k}} + 2)$ columns accepted by $\mathcal{A}$.

\begin{proof}
Let $H = k^{2^{2k}}$. Consider a two-dimensional word $z \in L(\mathcal{A})$ of dimension $|w| \times n_{1}$, where
\begin{equation}\label{eq:n1}
n_{1} > (|w| + 3) \cdot (H + 2).
\end{equation}
Let $C_{z}$ be an accepting nondeterministic computation of $\mathcal{A}$ on input word $z$. Without loss of generality, we may assume that $C_{z}$ accepts at the bottom border of $z$.

By the inequality in Equation~\ref{eq:n1}, the input word $z$ must have $k^{2k} + 1$ consecutive columns such that the computation $C_{z}$ does not make a downward move in any such column. Furthermore, we may assume that these consecutive columns do not include either the first $H$ columns or the last $H$ columns of $z$. That is, there exists $H \leq j \leq (n_{1} - 2H)$ such that the computation $C_{z}$ does not make a downward move in any of the subword occurrences
\begin{equation*}
\rowsubword_{i}[j, H], \ \ i = \{1, \dots, |w|\}.
\end{equation*}

Let $Q$ be the set of states of $\mathcal{A}$ and define $\overline{Q} = \{\overline{q} \mid q \in Q\}$ to be a disjoint copy of states in $Q$. For each column $x \in \{j, j + 1, \dots, j + H\}$, define a function $f_{x}: Q \to 2^{Q \cup \overline{Q}}$ by setting, for all $p \in Q$,
\begin{itemize}
\item $q \in f_{x}(p)$ if, for some $i$, the computation $C_{z}$ on the $i$th row in column $x$ and state $p$ exits the subword occurrence $\rowsubword_{i}[j, H]$ to the left in state $q$; and
\item $\overline{q} \in f_{x}(p)$ if, for some $i$, the computation $C_{z}$ on the $i$th row in column $x$ and state $p$ exits the subword occurrence $\rowsubword_{i}[j, H]$ to the right in state $q$.
\end{itemize}
Note that the computation of $C_{z}$ may visit the subword occurrence multiple times. By our definition, $q \in f_{x}(p)$ if, at some point, $C_{z}$ is in the $x$th column in state $p$ and, when $C_{z}$ next exits $\rowsubword_{i}[j, H]$, it exits to the left in state $q$.

Note also that the accepting computation must exit each subword occurrence $\rowsubword_{i}[j, H]$ either to the left or to the right since, by our choice of $j$, the computation $C_{z}$ makes no downward moves in any of the columns $j, \dots, (j + H)$.

Since the number of functions from $Q$ to $2^{Q \cup \overline{Q}}$ is $H = k^{2^{2k}}$, there exist columns $x_{1}$ and $x_{2}$, $j \leq x_{1} < x_{2} \leq (j + H)$, such that $f_{x_{1}} = f_{x_{2}}$. Moreover, since the computation $C_{z}$ makes no downward moves in any of the columns $j, \dots, (j + H)$, there exists an accepting computation of $\mathcal{A}$ on the two-dimensional word $z'$ obtained by removing the columns $x_{1}, \dots, (x_{2} - 1)$ from $z$.

The above observation relies on our earlier assumption that the designated columns $j, \dots, (j + H)$ are at distance at least $H$ from the left and right borders of the word. For example, consider a situation where $\overline{q} \in f_{x_{2}}(p)$; that is, where the computation starting in column $x_{2}$ and state $p$ exits the subword occurrence to the right in state $q$. When simulating the same computation on the modified word $z'$ starting in column $x_{1}$, the computation could, at some point, move to the left of column $j$. Since $j \geq H$, this guarantees that the computation would not reach the left border.

Altogether, the two-dimensional word $z'$ has $x_{2} - x_{1}$ fewer columns than the original word $z$. By repeated application of the previous argument, we see that $\mathcal{A}$ must accept a two-dimensional word of dimension $|w| \times n_{2}$, where $n_{2} \leq H$.
\end{proof}
\end{lemma}

An application of Lemma~\ref{lem:reduceworddimension} allows us to obtain our main space complexity result for column projections of languages recognized by unary three-way two-dimensional automata.

\begin{theorem}\label{thm:3WunaryautomataNLOGSPACE}
Let $\mathcal{A}$ be a unary three-way two-dimensional automaton. Then $\colp{L(\mathcal{A})} \in \NSPACE(O(\log(n)))$.

\begin{proof}
Suppose $\mathcal{A}$ has $k$ states. We describe the operation of a nondeterministic logspace Turing machine $\mathcal{M}$ recognizing $\colp{L(\mathcal{A})}$.

On input word $w$, $\mathcal{M}$ first writes to its work tape a binary representation of a nondeterministically-chosen natural number $n_{1} \leq (|w| + 3) \cdot (k^{2^{2k}} + 2)$. Since $k$ is constant, this binary representation can be written in space $O(\log(|w|))$.

The machine $\mathcal{M}$ then simulates a nondeterministic computation of $\mathcal{A}$ on a two-dimensional input word $z$ with $|w|$ rows and $n_{1}$ columns. The input head of $\mathcal{M}$ keeps track of the current row of $z$, while a binary counter stored on the work tape of $\mathcal{M}$ keeps track of the current column of $z$. The work tape also contains the originally-guessed value $n_{1}$ so that $\mathcal{M}$ is able to determine when its simulated computation encounters the right border of the input word.

By Lemma~\ref{lem:reduceworddimension}, we know that if $w \in \colp{L(\mathcal{A})}$, then $w$ must be a column projection of a two-dimensional word with at most $(|w| + 3) \cdot (k^{2^{2k}} + 2)$ columns that is accepted by $\mathcal{A}$.
\end{proof}
\end{theorem}

Since the language class \CSL\ coincides with the space complexity class $\NSPACE(O(n))$ \cite{Kuroda1965ClassesOfLanguages}, one consequence of Corollary~\ref{cor:4WautomataCSL} is that the row and column projections of languages recognized by four-way two-dimensional automata consist exactly of languages in $\NSPACE(O(n))$. Theorem~\ref{thm:3WunaryautomataNLOGSPACE} gives a significantly improved space complexity upper bound for column projections of languages recognized by unary three-way two-dimensional automata.

%%%%%%

\section{State Complexity}\label{sec:statecomplexity}

Since projections of languages in $L_{\TDFATWOW}$ and $L_{\TNFATWOW}$ are known to be always regular, it is possible to consider questions of state complexity involving these projection languages.

Although they seem never to have appeared anywhere in the literature, it is straightforward to prove the following closure results for Boolean operations over two-way two-dimensional automata.

\begin{lemma}\label{lem:2DFA2Wunionintersection}
The class $L_{\TDFATWOW}$ is not closed under union or intersection.

\begin{proof}
Take $\Sigma = \{\texttt{0}, \texttt{1}\}$. Define two languages $L_{1}$ and $L_{2}$ as follows:
\begin{align*}
L_{1}	&= \{w \in \Sigma^{2 \times 2} \mid w[0,0] = \texttt{1}, w[0,1] = \texttt{1}\}; \\
L_{2}	&= \{w \in \Sigma^{2 \times 2} \mid w[0,0] = \texttt{1}, w[1,0] = \texttt{1}\}.
\end{align*}
That is, $L_{1}$ is the language of $2 \times 2$ two-dimensional words with two \texttt{1}s in the first two positions of its first row, and $L_{2}$ is the language of $2 \times 2$ two-dimensional words with two \texttt{1}s in the first two positions of its first column.

Clearly, a deterministic two-way two-dimensional automaton can recognize words in $L_{1}$ by scanning the symbols at positions $(0,0)$ and $(0,1)$ and verifying that they are both \texttt{1}s. The same model can recognize words from $L_{2}$ in a similar manner.

However, no deterministic two-way two-dimensional automaton can recognize the language $L_{1} \cup L_{2}$.  Suppose such an automaton $\mathcal{A}$ can recognize the language $L_{1} \cup L_{2}$, and consider the computation of $\mathcal{A}$ on a $2 \times 2$ word $w \not\in L_{1} \cup L_{2}$ where $w[0,0] = \texttt{1}$ and all other symbols are \texttt{0}. This computation will reject $w$, but the input head will only scan one of the positions $(0,1)$ or $(1,0)$. Without loss of generality, suppose the unscanned position is $(1,0)$. Then the automaton would also reject a $2 \times 2$ word $w' \in L_{1} \cup L_{2}$ where $w'[0,0] = w'[1,0] = \texttt{1}$ and all other symbols are \texttt{0}.

For the same reason, no deterministic two-way two-dimensional automaton can recognize the language $L_{1} \cap L_{2}$. Consider a $2 \times 2$ word $x \in L_{1} \cap L_{2}$. Such a word has the symbol \texttt{1} at each of the positions $(0,0)$, $(0,1)$, and $(1,0)$, but no deterministic two-way two-dimensional automaton can scan all three of these positions. Given a $2 \times 2$ word $x' \not\in L_{1} \cap L_{2}$, where $x'[0,0] = x'[0,1] = \texttt{1}$ and $x'[1,0] = \texttt{0}$, a deterministic two-way two-dimensional automaton making one rightward move cannot distinguish between $x$ and $x'$.
\end{proof}
\end{lemma}

\begin{lemma}\label{lem:2NFA2Wunionintersection}
The class $L_{\TNFATWOW}$ is closed under union, but is not closed under intersection or complement.

\begin{proof}
Let $\mathcal{A}$ be a nondeterministic two-way two-dimensional automaton, and let $L$ and $L'$ be languages recognized by this model. Then $\mathcal{A}$ can recognize the language $L \cup L'$ by making a nondeterministic selection between $L$ and $L'$ at the beginning of its computation, and checking whether its input word $w$ belongs to the chosen language.

Intersection is not closed for this model for the same reason as given in the proof of Lemma~\ref{lem:2DFA2Wunionintersection}.

As a consequence of this model being closed under union but not intersection, we necessarily cannot have closure under complement.
\end{proof}

\end{lemma}

Moreover, the present authors previously investigated closure properties of concatenation operations over two-way two-dimensional automata~\cite{SmithSalomaa20202DConcatenationPreprint}. In this section, therefore, we will focus on the state complexity of projections of union and concatenation operations for nondeterministic two-way two-dimensional automata.

%%%

\subsection{Union of \TNFATWOW\ Languages}\label{subsec:statecomplexityunion2W}

Before we proceed, we require a slight modification to the definition of a two-way two-dimensional automaton that we introduced in Section~\ref{sec:prelim}. 
For the remainder of this section, when we refer to a ``two-way two-dimensional automaton", we use the following definition.

\begin{definition}[IBR-accepting two-way two-dimensional automaton]
An IBR-accepting two-way two-dimensional automaton $\mathcal{A}$ is a tuple $(Q, \Sigma, \delta, q_{0}, q_{\text{accept}})$ as in Definition~\ref{def:2DFA}, where, 
when the input head reads a boundary marker \# for the first time, $\mathcal{A}$ either enters $q_{\text{accept}}$ in the next transition or the transition is undefined.
\end{definition}

The abbreviation ``IBR-accepting" refers to the automaton ``immediately-bottom-right accepting", or accepting only once the input head reaches the bottom or right border of the input word. The two-way model is the only model for which we can make this modification; neither three- nor four-way models can be made to halt immediately upon reading a boundary marker.

\begin{remark}
The accepting state of an IBR-accepting two-way two-dimensional automaton, $q_{\text{accept}}$, is a ``dummy" state used only as the target of accepting transitions on the boundary symbol \#. 
Thus, by the ``size" of such an automaton $\mathcal{A}$ we mean the size of the set $Q - \{q_{\text{accept}}\}$. This convention ensures that an IBR-accepting two-way two-dimensional automaton recognizing single-row words has the same size as the corresponding one-dimensional automaton accepting the same string language.
\end{remark}

The following result shows that we may convert between the usual and IBR-accepting types of two-way two-dimensional automata without incurring a penalty on the number of states.

\begin{proposition}[\cite{SmithSalomaa20202DConcatenationPreprint}]\label{prop:IBRacceptingstatecomplexity}
Given a two-way two-dimensional automaton $\mathcal{A}$ with $n$ states, there exists an equivalent IBR-accepting two-way two-dimensional automaton $\mathcal{A}'$ with $n$ states.

\begin{proof}[Proof Sketch]
If $\mathcal{A}$ reads a boundary marker, then its input head can never reenter the input word. After reading a boundary marker in state $q_{i}$, say, we can decide whether $q_{\text{accept}}$ is reachable from $q_{i}$ after following some number of transitions of $\mathcal{A}$. Thus, we may take $\mathcal{A}'$ to be the same as $\mathcal{A}$ where the transition upon reading the boundary marker \# goes directly to $q_{\text{accept}}$ if that state is reachable or is undefined otherwise.
\end{proof}
\end{proposition}

Using a construction from a previous paper investigating projections of nondeterministic two-way two-dimensional automaton languages~\cite{SmithSalomaa20192D3WProperties}, we may obtain an upper bound on the nondeterministic state complexity of projection languages for this model.

\begin{proposition}\label{prop:2NFA2Wrowprojectionupper}
Let $\mathcal{A}$ be a nondeterministic two-way two-dimensional automaton with $n$ states. Then both $\rowp{L(\mathcal{A})}$ and $\colp{L(\mathcal{A})}$ are recognized by a nondeterministic one-dimensional automaton with $2n$ states.

\begin{proof}[Proof Sketch]
Given a nondeterministic two-way two-dimensional automaton $\mathcal{A}$, we may construct a nondeterministic one-dimensional automaton $\mathcal{B}$ recognizing the language $\rowp{L(\mathcal{A})}$ that simulates rightward moves of $\mathcal{A}$ and keeps track of whether a downward move is made during the computation of $\mathcal{A}$. We may remember downward moves by doubling the number of states of $\mathcal{A}$. Using an analogous construction, we obtain the same result for $\colp{L(\mathcal{A})}$.
\end{proof}
\end{proposition}

We can show that the following lower bound applies for the same model.

\begin{lemma}\label{lem:2NFA2Wrowprojectionlower}
There exists a nondeterministic two-way two-dimensional automaton $\mathcal{A}$ with $n$ states such that any nondeterministic one-dimensional automaton recognizing $\rowp{L(\mathcal{A})}$ requires at least $2n - 1$ states.

\begin{proof}
Define $\mathcal{A}$ as follows: the alphabet is $\Sigma = \{\texttt{0}, \texttt{1}\}$, the set of states is $Q = \{q_{0}, q_{1}, \dots, q_{n-1}\}$ (and additionally $q_{\text{accept}}$), the initial state is $q_{0}$, the accepting state is $q_{\text{accept}}$, and the transition function $\delta$ consists of the following:
\begin{itemize}
	\item $\delta(q_{i}, \texttt{0}) = (q_{i+1}, R)$ for all $0 \leq i \leq n-2$;
	\item $\delta(q_{n-1}, \texttt{0}) = \{(q_{0}, R), (q_{n-1}, D)\}$; and
	\item $\delta(q_{0}, \#) = (q_{\text{accept}}, N)$.
\end{itemize}
Each rightward-moving transition counts modulo $n$, and the only downward-moving transition occurs in a column position congruent to $-1 \bmod n$. Moreover, the downward-moving transition does not change the state (i.e., the column count is preserved). Note also that $\mathcal{A}$ makes no transitions upon reading the symbol \texttt{1}; this is because, after reading $n - 1$ copies of \texttt{0} and making a downward move, the first row can contain any symbols after that column position so long as the number of total columns remains a multiple of $n$. Combining these observations, we see that the row projection of $L(\mathcal{A})$ is
\begin{equation*}
L_{\text{pr}} = \texttt{0}^{n-1} (\texttt{0} + \texttt{1}) ((\texttt{0} + \texttt{1})^{n})^{*} + \epsilon.
\end{equation*}

To show that the nondeterministic state complexity of $L_{\text{pr}}$ is at least $2n - 1$, we use the following extended fooling set \cite{HolzerKutrib2003NondeterministicDescriptionalComplexity}:
\begin{equation*}
S = \{(x, y) \mid xy = \texttt{0}^{n - 1} \texttt{1}^{n + 1}, |y| \geq 2\}.
\end{equation*}
The set $S$ contains $2n - 1$ elements and, by its definition, for any pair $(x, y) \in S$, $xy = \texttt{0}^{n - 1}\texttt{1}^{n + 1} \in L_{\text{pr}}$.

Consider two distinct pairs $(x, y)$ and $(x', y')$. Without loss of generality, assume $x$ is a proper prefix of $x'$. 
If $|x'| - |x| \neq n$, then $|xy'|$ is not a multiple of $n$, and $xy' \not\in L_{\text{pr}}$. 
Otherwise, $|x'| - |x| = n$. In this case, since $|x'y'| = 2n$ and $|y'| \geq 2$, we have that $|x'| \leq 2n - 2$. Thus, $|x| \leq n - 2$, and so $x = \texttt{0}^{i}$ for some $0 \leq i \leq n - 2$. However, this means that $xy' \not\in L_{\text{pr}}$, because in this case $y'$ consists only of the symbol \texttt{1}.
\end{proof}
\end{lemma}

Using the previous results, we can obtain a state complexity bound for the projection of the union of two languages recognized by nondeterministic two-way two-dimensional automata.

\begin{theorem}\label{thm:2NFA2Wstatecomplexityunion}
(i) If $\mathcal{A}$ and $\mathcal{B}$ are nondeterministic two-way two-dimensional automata with $m$ and $n$ states, respectively, then $\rowp{L(\mathcal{A}) \cup L(\mathcal{B})}$ is recognized by a nondeterministic one-dimensional automaton with $2(m + n + 1)$ states.

(ii) There exist nondeterministic two-way two-dimensional automata $\mathcal{A}$ and $\mathcal{B}$ with $n$ and $m$ states, respectively, such that any nondeterministic one-dimensional automaton recognizing $\rowp{L(\mathcal{A}) \cup L(\mathcal{B})}$ requires at least $2(m + n - 1)$ states.

\begin{proof}
We prove (i) by construction. Without loss of generality, assume the state sets of $\mathcal{A}$ and $\mathcal{B}$ are disjoint. Then a nondeterministic two-way two-dimensional automaton $\mathcal{C}$ recognizing the language $L(\mathcal{A}) \cup L(\mathcal{B})$ can be constructed in the following way:
\begin{itemize}
\item The state set of $\mathcal{C}$ contains all  
non-accepting states of $\mathcal{A}$ and $\mathcal{B}$.
\item There exists a new initial state $q_{0_{\mathcal{C}}}$ that simulates outgoing transitions from the original initial states $q_{0_{\mathcal{A}}}$ and $q_{0_{\mathcal{B}}}$. 
\item There exists a joint accepting state $q_{\text{accept}_{\mathcal{C}}}$. 
\item The transition function $\delta_{\mathcal{C}}$ includes all transitions of $\delta_{\mathcal{A}}$ and $\delta_{\mathcal{B}}$.
\end{itemize}
Since the state sets of $\mathcal{A}$ and $\mathcal{B}$ are disjoint, $\mathcal{C}$ accepts some input $w$ if and only if $w$ is accepted by either $\mathcal{A}$ or $\mathcal{B}$. By our construction, $\mathcal{C}$ consists of $m + n + 1$ states. Since $\mathcal{C}$ must remember whether or not a downward move is made during the computation of $\mathcal{A}$ (or, similarly, during the computation of $\mathcal{B}$), we must double the number of states of $\mathcal{C}$; this is essentially the same construction as that used by Proposition~\ref{prop:2NFA2Wrowprojectionupper}. Therefore, 
$\rowp{L(\mathcal{A}) \cup L(\mathcal{B})}$ is recognized by a nondeterministic one-dimensional automaton with $2(m + n + 1)$ states.

We now prove (ii). Let $\mathcal{A}$ be the $n$-state nondeterministic two-way two-dimensional automaton from the proof of Lemma~\ref{lem:2NFA2Wrowprojectionlower}, and let $\mathcal{B}$ be a ``copy" of $\mathcal{A}$ with $m$ states; specifically, $\mathcal{B}$ is an automaton of the same type as $\mathcal{A}$ over the alphabet $\Sigma' = \{\texttt{2}, \texttt{3}\}$ where the set of states is $Q' = \{q_{0}, q_{1}, \dots, q_{m-1}\}$ (and additionally $q_{\text{accept}}$); the transition function $\delta'$ is identical to $\delta$ with \texttt{0}, \texttt{1}, and $n$ replaced by \texttt{2}, \texttt{3}, and $m$, respectively; and all other aspects are the same.

Let $L_{\text{pr}} = \rowp{L(\mathcal{A}) \cup L(\mathcal{B})}$. A fooling set for $L_{\text{pr}}$ is
\begin{equation*}
S = \{(x, y) \mid xy = \texttt{0}^{n - 1} \texttt{1}^{n + 1}, |y| \geq 2\} \cup \{(x, y) \mid xy = \texttt{2}^{m - 1} \texttt{3}^{m + 1}, |y| \geq 2\}.
\end{equation*}
The set $S$ contains $2(m + n - 1)$ elements. Moreover, $S$ is clearly a fooling set, since mixing a pair over the alphabet $\{\texttt{0}, \texttt{1}\}$ and a pair over the alphabet $\{\texttt{2}, \texttt{3}\}$ always produces strings not in $L_{\text{pr}}$.
\end{proof}
\end{theorem}

Since two-way two-dimensional automata operate symmetrically with respect to rows and columns, there also exist nondeterministic state complexity bounds for column projections analogous to those established in Theorem~\ref{thm:2NFA2Wstatecomplexityunion}.

%%%

\subsection{Diagonal Concatenation of \TNFATWOW\ Languages}\label{subsec:statecomplexityconcat2W}

Given two-dimensional words $w$ and $v$ of dimension $m \times n$ and $m' \times n'$ respectively, the diagonal concatenation of $w$ and $v$, denoted $w \oslash v$, produces a two-dimensional language consisting of words of dimension $(m + m') \times (n + n')$ where $w$ is in the top-left corner, $v$ is in the bottom-right corner, and 
words $x \in \Sigma^{m \times n'}$ and $y \in \Sigma^{m' \times n}$ are placed in the ``top-right" and ``bottom-left" corners of $w \oslash v$, respectively. We assume that the symbols in $x$ and $y$ come from the same alphabet $\Sigma$ as the symbols in $w$ and $v$. 
The diagonal concatenation language is formed by adding to the corners all possible words $x$ and $y$ over $\Sigma$. 
An example word from such a language is depicted in Figure~\ref{fig:2Ddiagonalconcatenation}.

\begin{figure}[t]
\[\arraycolsep=1.4pt\def\arraystretch{0.8}
w \oslash v = 
\begin{array}{cccccccc}
\#		& \#		& 		& \#			& \#		& 		& \# 		& \# \\
\#		& w_{1,1}	& \cdots	& w_{1,n} 	& x_{1,1}		& \cdots	& x_{1,n'}	& \# \\
		& \vdots	&		& \vdots	& \vdots		& 		& \vdots	& \\
\#		& w_{m,1}	& \cdots	& w_{m,n}	& x_{m,1}		& \cdots	& x_{m,n'}	& \# \\
\#		& y_{1,1}	& \cdots	& y_{1,n}	& v_{1,1}		& \cdots	& v_{1,n'}	& \# \\
		& \vdots	&		& \vdots	& \vdots		&		& \vdots	& \\
\#		& y_{m',1}	& \cdots	& y_{m',n}	& v_{m',1}		& \cdots	& v_{m',n'}& \# \\
\#		& \#		& 		& \#			& \#		& 		& \# 		& \#
\end{array}
\]
\caption{Diagonal concatenation of two-dimensional words}
\label{fig:2Ddiagonalconcatenation}
\end{figure}

Nondeterministic two-way two-dimensional automata are known to be closed under diagonal concatenation over a general alphabet and, moreover, this is the only 
concatenation operation under which two-way two-dimensional automaton languages over general alphabets are closed 
\cite{SmithSalomaa20202DConcatenationPreprint}. Thus, the natural question arises: given a pair of nondeterministic two-way two-dimensional automata $\mathcal{A}$ and $\mathcal{B}$ recognizing languages $L(\mathcal{A})$ and $L(\mathcal{B})$, respectively, how large must such an automaton be to recognize $\rowp{L(\mathcal{A}) \oslash L(\mathcal{B})}$?

We begin by making an elementary observation. In one dimension, an $\epsilon$-NFA extends an ordinary NFA by allowing $\epsilon$-transitions; i.e., ``stay-in-place" moves. The following result is well-known:
\begin{lemma}[Wood \cite{Wood1987TheoryComputation}]\label{lem:epsilonNFA}
Any $n$-state $\epsilon$-NFA has an equivalent $n$-state NFA without $\epsilon$-transitions.
\end{lemma}

Moreover, for a pair of nondeterministic one-dimensional automata with $m'$ and $n'$ states recognizing languages $L_{1}$ and $L_{2}$, respectively, a total of $m' + n'$ states are necessary and sufficient to recognize the concatenation language $L_{1} \cdot L_{2}$ in the general alphabet case, while $m' + n' - 1$ states are necessary in the unary case \cite{HolzerKutrib2003NondeterministicDescriptionalComplexity}.

\begin{theorem}\label{thm:2NFA2Wstatecomplexitydiagconcat}
(i) If $\mathcal{A}$ and $\mathcal{B}$ are nondeterministic two-way two-dimensional automata with $m$ and $n$ states, respectively, then $\rowp{L(\mathcal{A}) \oslash L(\mathcal{B})}$ is recognized by a nondeterministic one-dimensional automaton with $2m + n$ states.

(ii) There exist nondeterministic two-way two-dimensional automata $\mathcal{A}$ and $\mathcal{B}$ with $m$ and $n$ states, respectively, such that any nondeterministic one-dimensional automaton recognizing $\rowp{L(\mathcal{A}) \oslash L(\mathcal{B})}$ requires at least $m + n - 1$ states.

\begin{proof}
We prove (i) by constructing a nondeterministic one-dimensional automaton $\mathcal{C}$ to recognize the language $\rowp{L(\mathcal{A}) \oslash L(\mathcal{B})}$. The following procedure allows $\mathcal{C}$ to simulate the computation of $\mathcal{A}$ and $\mathcal{B}$ on a word in the language $L(\mathcal{A}) \oslash L(\mathcal{B})$:
\begin{enumerate}
\item The input head of $\mathcal{C}$ begins by simulating rightward moves of the input head of $\mathcal{A}$. If the input head of $\mathcal{A}$ makes a downward move, $\mathcal{C}$ remembers that a downward move occurred and replaces it with a ``stay-in-place" move.
\item At some point during its computation, $\mathcal{C}$ nondeterministically switches to simulating moves of $\mathcal{B}$. Again, the input head of $\mathcal{C}$ only simulates rightward moves, and replaces downward moves with ``stay-in-place" moves.
\end{enumerate}

By Lemma~\ref{lem:epsilonNFA}, ``stay-in-place" moves can be used without affecting the number of states. However, by a construction similar to that used in Proposition~\ref{prop:2NFA2Wrowprojectionupper}, the requirement in Step 1 to remember whether a downward move occurred doubles the number of states needed to simulate the computation of $\mathcal{A}$. Remembering downward moves is not required when simulating the computation of $\mathcal{B}$. Furthermore, in Step 2, the input head of $\mathcal{C}$ ignores the alphabet symbols it is reading. Since the simulation only needs to check that $\mathcal{B}$ accepts a two-dimensional word with the correct number of columns, the exact symbols being read at this stage may be ignored.

If the computation of $\mathcal{C}$ accepts, then the computation of $\mathcal{A}$ and $\mathcal{B}$ must have also accepted, and therefore $\mathcal{C}$ recognizes words in the language $\rowp{L(\mathcal{A}) \oslash L(\mathcal{B})}$. Moreover, $2m + n$ states are sufficient for $\mathcal{C}$ to perform its computation in this way.

We now prove (ii). Let $\mathcal{A}'$ (respectively, $\mathcal{B}'$) be an $m$-state (respectively, $n$-state) unary nondeterministic one-dimensional automaton such that the concatenation of $L(\mathcal{A}')$ and $L(\mathcal{B}')$ requires $m + n - 1$ states \cite{HolzerKutrib2003NondeterministicDescriptionalComplexity}. The language $L(\mathcal{A}')$ can be recognized by an $m$-state nondeterministic two-way two-dimensional automaton $\mathcal{A}$ that recognizes words consisting of one row. Similarly, $L(\mathcal{B}')$ can be recognized by an $n$-state nondeterministic two-way two-dimensional automaton $\mathcal{B}$. In this case, the languages $\rowp{L(\mathcal{A}) \oslash L(\mathcal{B})}$ and $L(\mathcal{A}') \cdot L(\mathcal{B}')$ are equal. It follows that $m + n - 1$ states are necessary for any nondeterministic one-dimensional automaton to recognize $\rowp{L(\mathcal{A}) \oslash L(\mathcal{B})}$.
\end{proof}
\end{theorem}

Again, there exist nondeterministic state complexity bounds for column projections analogous to those established in Theorem~\ref{thm:2NFA2Wstatecomplexitydiagconcat}.

%%%%%%

\section{Conclusion}\label{sec:conclusion}

In this paper, we established results linking one-dimensional language classes to two-dimensional projection languages; namely, that both the row and column projections of languages $L \in L_{\TDFA}$ or $L_{\TNFA}$ are exactly context-sensitive. This improves on the 
previously-known non-context-free lower bound, which remains for other two-dimensional automaton models.

We also proved space complexity results for projection languages. While both the row and column projections of languages $L \in L_{\TDFA}$ or $L_{\TNFA}$ belong to the class $\NSPACE(O(n))$, the column projection of languages $L \in L_{\TDFATWOS}$ or $L_{\TNFATWOS}$ belongs to the class $\NSPACE(O(\log(n)))$.

Finally, we investigated the state complexity of projection languages. We showed that, given a pair of nondeterministic two-way two-dimensional automata $\mathcal{A}$ and $\mathcal{B}$ with $m$ and $n$ states, respectively, between $2(m + n - 1)$ and $2(m + n + 1)$ states are needed to recognize $\rowp{L(\mathcal{A}) \cup L(\mathcal{B})}$ and between $m + n - 1$ and $2m + n$ states are needed to recognize $\rowp{L(\mathcal{A}) \oslash L(\mathcal{B})}$. 
These bounds apply also to the column projections of such languages.

We conclude by giving a selection of open problems arising from work done in this paper.
\begin{enumerate}
\item\label{itm:prob1} Which class of one-dimensional languages corresponds to $L_{\TDFATW}$/$L_{\TNFATW}$ (or their unary equivalents) under the operation of column projection?
\item\label{itm:prob2} Which class of one-dimensional languages corresponds to $L_{\TDFAOS}$/$L_{\TNFAOS}$ under the operations of row and column projection?
\item\label{itm:prob3} If a two-dimensional automaton $\mathcal{A}$ with $n$ states recognizes a language $L$, how many states are necessary/sufficient for a one-dimensional automaton $\mathcal{A}'$ to recognize the language $\rowp{L}$/$\colp{L}$?
\end{enumerate}
Problems~\ref{itm:prob1} and \ref{itm:prob2} are likely difficult; it may be more reasonable to obtain an improved upper bound on the related question of space complexity for problem~\ref{itm:prob2}, say $\DSPACE(O(n))$. Moreover, for problem~\ref{itm:prob3}, we can obtain a trivial lower bound of $n$ states by constructing an $n$-state nondeterministic three/four-way two-dimensional automaton $\mathcal{A}$ that accepts only words of dimension $1 \times k$, $k \geq 1$, and taking $\mathcal{A}'$ to be the minimal nondeterministic two-way one-dimensional automaton recognizing the language $\rowp{L(\mathcal{A})}$.

%%%%%%

\bibliographystyle{plain}
\bibliography{./References.bib}

\begin{thebibliography}{10}

\bibitem{Anselmo2010DeterministicUnambiguousFamilies}
Marcella Anselmo, Dora Giammarresi, and Maria Madonia.
\newblock Deterministic and unambiguous families within recognizable
  two-dimensional languages.
\newblock {\em Fundamenta Informaticae}, 98(2--3):143--166, 2010.

\bibitem{Anselmo2011ClassificationTilingRecognizable}
Marcella Anselmo, Dora Giammarresi, and Maria Madonia.
\newblock Classification of string languages via tiling recognizable picture
  languages.
\newblock In A.-H. Dediu, S.~Inenaga, and C.~Mart{\'i}n-Vide, editors, {\em
  Proceedings of the 5th International Conference on Language and Automata
  Theory and Applications ({LATA} 2011)}, volume 6638 of {\em Lecture Notes in
  Computer Science}, pages 105--116, Berlin Heidelberg, 2011. Springer-Verlag.

\bibitem{BlumHewitt19672DAutomata}
Manuel Blum and Carl Hewitt.
\newblock Automata on a 2-dimensional tape.
\newblock In R.~E. Miller, editor, {\em Proceedings of the 8th Annual Symposium
  on Switching and Automata Theory ({SWAT} 1967)}, pages 155--160, 1967.

\bibitem{GiammarresiRestivo1992RecognizablePictureLanguages}
Dora Giammarresi and Antonio Restivo.
\newblock Recognizable picture languages.
\newblock {\em International Journal of Pattern Recognition and Artificial
  Intelligence}, 6(2--3):241--256, 1992.

\bibitem{GiammarresiRestivo19972DLanguages}
Dora Giammarresi and Antonio Restivo.
\newblock Two-dimensional languages.
\newblock In G.~Rozenberg and A.~Salomaa, editors, {\em Handbook of Formal
  Languages}, volume~3, pages 215--267. Springer-Verlag, Berlin Heidelberg,
  1997.

\bibitem{Hartmanis1968RecognitionPrimesAutomata}
Juris Hartmanis and Herbert Shank.
\newblock On the recognition of primes by automata.
\newblock {\em Journal of the {ACM}}, 15(3):382--389, 1968.

\bibitem{HolzerKutrib2003NondeterministicDescriptionalComplexity}
Markus Holzer and Martin Kutrib.
\newblock Nondeterministic descriptional complexity of regular languages.
\newblock {\em International Journal of Foundations of Computer Science},
  14(6):1087--1102, 2003.

\bibitem{HopcroftUllman1979IntroAutomataTheory}
John~E. Hopcroft and Jeffrey~D. Ullman.
\newblock {\em Introduction to Automata Theory, Languages, and Computation}.
\newblock Addison-Wesley, Reading, 1979.

\bibitem{Inoue19912DAutomataSurvey}
Katsushi Inoue and Itsuo Takanami.
\newblock A survey of two-dimensional automata theory.
\newblock {\em Information Sciences}, 55(1--3):99--121, 1991.

\bibitem{Inoue1992CharacterizationRecognizable}
Katsushi Inoue and Itsuo Takanami.
\newblock A characterization of recognizable picture languages.
\newblock In A.~Nakamura, M.~Nivat, M.~Saoudi, P.~S.~P. Wang, and K.~Inoue,
  editors, {\em Proceedings of the 2nd International Conference on Parallel
  Image Analysis ({ICPIA} 1992)}, volume 654 of {\em Lecture Notes in Computer
  Science}, pages 133--143, Berlin Heidelberg, 1992. Springer-Verlag.

\bibitem{KariMoore2004RectanglesAndSquares}
Jarkko Kari and Cristopher Moore.
\newblock Rectangles and squares recognized by two-dimensional automata.
\newblock In J.~Karhum{\"a}ki, H.~Maurer, G.~Paun, and G.~Rozenberg, editors,
  {\em Theory is Forever: Essays Dedicated to {A}rto {S}alomaa on the Occasion
  of His 70th Birthday}, volume 3113 of {\em Lecture Notes in Computer
  Science}, pages 134--144, Berlin Heidelberg, 2004. Springer-Verlag.

\bibitem{Kuroda1965ClassesOfLanguages}
Sige-Yuki Kuroda.
\newblock Classes of languages and linear-bounded automata.
\newblock {\em Information and Control}, 7(2):207--223, 1965.

\bibitem{Latteux1997ContextSensitiveRecognizable}
Michel Latteux and David Simplot.
\newblock Context-sensitive string languages and recognizable picture
  languages.
\newblock {\em Information and Computation}, 138(2):160--169, 1997.

\bibitem{Morita20042DLanguages}
Kenichi Morita.
\newblock Two-dimensional languages.
\newblock In C.~Mart{\'i}n-Vide, V.~Mitrana, and G.~P{\u a}un, editors, {\em
  Formal Languages and Applications}, volume 148 of {\em Studies in Fuzziness
  and Soft Computing}, pages 427--437. Springer-Verlag, Berlin Heidelberg,
  2004.

\bibitem{Rosenfeld1979PictureLanguages}
Azriel Rosenfeld.
\newblock {\em Picture Languages: Formal Models for Picture Recognition}.
\newblock Computer Science and Applied Mathematics. Academic Press, New York,
  1979.

\bibitem{Salomaa1969TheoryOfAutomata}
Arto Salomaa.
\newblock {\em Theory of Automata}, volume 100 of {\em International Series of
  Monographs in Pure and Applied Mathematics}.
\newblock Pergamon Press, Oxford, 1969.

\bibitem{Salomaa1973FormalLanguages}
Arto Salomaa.
\newblock {\em Formal Languages}.
\newblock Academic Press, New York, 1973.

\bibitem{Shepherdson1959TwoWayReduction}
John~C. Shepherdson.
\newblock The reduction of two-way automata to one-way automata.
\newblock {\em {IBM} Journal of Research and Development}, 3(2):198--200, 1959.

\bibitem{Sipser1997Computation}
Michael Sipser.
\newblock {\em Introduction to the Theory of Computation}.
\newblock {PWS} Publishing Company, Boston, 1997.

\bibitem{Smith2019TwoDimensionalAutomata}
Taylor~J. Smith.
\newblock Two-dimensional automata.
\newblock Technical report 2019-637, Queen's University, Kingston, 2019.

\bibitem{SmithSalomaa20192D3WProperties}
Taylor~J. Smith and Kai Salomaa.
\newblock Decision problems for restricted variants of two-dimensional
  automata.
\newblock In M.~Hospod{\'a}r and G.~Jir{\'a}skov{\'a}, editors, {\em
  Proceedings of the 24th International Conference on Implementation and
  Application of Automata ({CIAA} 2019)}, volume 11601 of {\em Lecture Notes in
  Computer Science}, pages 222--234, Berlin Heidelberg, 2019. Springer-Verlag.

\bibitem{SmithSalomaa20202DConcatenationPreprint}
Taylor~J. Smith and Kai Salomaa.
\newblock Concatenation operations and restricted variants of two-dimensional
  automata, 2020.
\newblock arXiv:2008.11164.

\bibitem{Wood1987TheoryComputation}
Derick Wood.
\newblock {\em Theory of Computation}.
\newblock Harper \& Row Computer Science and Technology Series. Harper \& Row,
  New York, 1987.

\end{thebibliography}

%%%%%%

\end{document}